\documentclass{article}

\usepackage{arxiv}

\usepackage[english]{babel}
\usepackage[T1]{fontenc}
\usepackage{amsmath}
\usepackage{amsfonts}
\usepackage{amssymb}
\usepackage{url}

\usepackage{graphicx}
\usepackage[font=small,margin=0.9cm]{caption}
\usepackage[usenames,dvipsnames]{xcolor}
\usepackage{mathtools,mathrsfs}
\usepackage{nicefrac}      
\usepackage{subfig}
\usepackage{booktabs}
\usepackage{tabularx,multirow}
\usepackage{microtype}

\newcommand{\numberset}{\mathbb}

\newcommand{\R}{\numberset{R}}

\title{What can we learn from functional clustering of mortality data?\\
An application to HMD data}
\author{Ainhoa-Elena Léger \\
email: ainhoaelena.leger@gmail.com \\
\And Stefano Mazzuco\\
Department of Statistical Sciences, University of Padova\\
Via Cesare Battisti 241, 35121 Padova, Italy\\
email: mazzuco@stat.unipd.it
}

\begin{document}
\maketitle

\begin{abstract}
In most cases, mortality is analysed considering summary indicators (e.~g. $e_0$ or $e^{\dagger}_0$) that either focus on a specfic mortality component or pool all component-specific information in one measure. This can be a limitation, when we are interested to analyse the global evolution of mortality patterns without loosing sight of specific components evolution. The paper analyses whether there are different patterns of mortality decline among developed countries, identifying the role played by all the mortality components. We implement a cluster analysis using a Functional Data Analysis (FDA) approach, which allows us to consider age-specific mortality rather than summary measures as it analyses curves rather than scalar data. Combined with a Functional Principal Component Analysis (PCA) method it can identify what part of the curves (mortality components) is responsible for assigning one country to a specific cluster. FDA clustering is applied to 32 countries of Human Mortality Database and years 1960--2010. The results show that the evolutions of developed countries follow the same pattern (with different timing): (1) a reduction of infant mortality, (2) an increase of premature mortality, (3) a shift and compression of deaths. Some countries are following this scheme and recovering the gap with precursors, others do not show signs of recovery. Eastern Europe countries are still at stage (2) and it is not clear if and when they will enter into phase (3). All the country differences relates the different timing with which countries undergo the stages identified by clusters. The cluster analysis based on FDA allows therefore a comprehensive understanding of the patterns of mortality decline for considered countries.
\end{abstract}


\section{Introduction}
Mortality modeling and prediction of its future trends can provide fundamental answers to several key questions related to longevity ranking and demographic sustainability, among others. However, in most of the cases, this is done by focusing on summary measures like life expectancy at birth $e_0$ or life disparity $e^{\dagger}_0$. For example \cite{AminSteinmetz} link life expectancy with cardiovascular disease and cancer in US states finding spatial clusters with higher values of $e_0$. In 1992, \cite{LeeCarter} developed a model to forecast mortality based on a matrix of the logged death rates by age and time decomposed into a single time-index and an age-pattern of mortality changes. This model is the same that lead \cite{Tuljapurkar} to suggest the existence of an universal pattern of mortality decline. Life expectancy at birth is also applied to evaluate the precision of mortality forecasts, even though \cite{Bohk-Ewald2017} has suggested that lifespan disparity could also be used. Lifespan disparity has also been advocated as an useful indicator to analyse the mortality evolution of countries (\cite{Vaupele000128}). In other cases, scholars focus on specific components of mortality, disregarding the global pattern. For instance, \cite{Medford2019}, analyse lifespan after age 100 in Sweden and Denmark to show that Danish centenarians lifespans have been lengthening, but not those of Swedish. As another  example, \cite{Zanotto2016} focus their analysis on premature mortality. Therefore, it looks like that analysing or predicting mortality evolution of one or more countries means choosing among focusing on global mortality or a specific component. In this work, we suggest to apply Functional Data Analysis (FDA) approach to mortality data. Such approach (\cite{Ramsay1}) is increasingly gaining ground among scholars interested to analyse curves rather than scalar data, and mortality profiles (e.g. in terms of age-specific rates) can be seen as curves over age that can be observed for every country and every year.\\
More specifically we propose a functional clustering of mortality profiles of several countries. \cite{HatzHab} already tried a clustering solution, using a fuzzy c-means cluster analysis based on the main time trends, which are estimated by means of a GLM model, supporting the idea of a single mortality pattern of mortality decline across subpopulations. However, other authors contrast this hypothesis. For instance, \cite{MCMICHAEL20041155} show there is an increased heterogeneity across countries, even though it should be noted that in their analysis both developed and poor countries are considered. We suggest that taking a functional perspective can be a more informative approach as it allows to cluster countries on the base of global mortality profiles without loosing sight of the role played by single components. Indeed, combining functional clustering with a Principal Component analysis permits to identify the components that determine the exclusion or inclusion of a country into a specific cluster. In this way, we can see whether countries mortality are evolving in the same way (i.e. following the same sequence of clusters) or different patterns are found.\\
The remainder of this paper is organised as follows: in the next section FDA-based clustering techniques are exposed in detail, while in section \ref{sec:Data} we explain our choice of data. Results of the analyses are reported in section \ref{sec:Results}, while section \ref{sec:Concl} concludes.
\section{Functional Clustering methods}
\label{sec:FDA}
Functional data analysis (FDA) deals with the analysis of data that are in the form of functions and extends the classical multivariate methods. The monographs on functional data by \cite{Ramsay1, ramsay2005analysis} developing methodology and applications and the book of \cite{ferraty2006nonparametric} on nonparametric models contain a review of the most recent contributions on this topic. Our work proposes to study mortality data using FDA, more specifically functional cluster analysis. Because of the nature of data itself, belonging to an infinite dimensional space, clustering functional data is generally a difficult task and several approaches have been proposed along the years. A review of clustering methods can be found in \cite{jacques2014functional}. We present in this section how to obtain a functional representation of data, the theory of functional principal component analysis and the major functional approaches for data clustering.
\subsection{Functional data} 
Functional approach to the analysis of data considers a collection of discrete observations $x(t_{1}),\dots,x(t_{N})$ at a finite set of instants $\{t_{1},...,t_{N}\}$ as coming from a continuous underlying function $x(t)$ defined on $t \in [0,T]$. Functional data consist then of a set of $n$ curves $x_{1}(t),\dots,x_{n}(t)$ defined on a common interval $[0,T]$, denoted in the case of observations at the same istant $t_{j}$ as \begin{equation} x_{i}(t_{j}), \quad t_{j} \in [0,T], \quad j=1,\dots,N, \quad i=1,2,\dots,n. \end{equation} The curves are assumed to be independent realizations drawn from the same continuous stochastic process $X(t)$ belonging to $L_{2}[0,T]$ space. \\
Because functional observations are supposed to belong to an infinite dimensional space, the first step in a FDA is often the reconstruction of the functional form of data. As a procedure of functional representation we approximate the function $X(t)$ by using a basis expansion of cubic \textit{B-splines} functions. Let us consider $p$ known basis functions $\psi(t)=(\psi_{1}(t),\dots,\psi_{p}(t))$, the basis expansion for $X(t)$ is \begin{equation} X(t)=\sum_{j=1}^{p} \gamma_{j}\psi_{j}(t), \end{equation} where $\gamma=(\gamma_{1},\dots,\gamma_{p})'$ are the basis function coefficients to be estimated. Resulting spline functions are piecewise polynomials defined into subintervals, with boundaries at points called breaks. Given an order $m$ ($\text{degree}-1$) and $L-2$ internal knots, there are $(L-2)+m$ basis functions. \\
The curves are observed with error, therefore \textit{B-splines} basis coefficients can be estimated by ordinary least squares method minimizing the sum of squared residuals. There exist many possible approaches to control the irregularity of the curve and obtain a better approximation. The one we use is a roughness penalty method, which modifies estimation criterion adding a penalty term, so the penalised sum of squared errors (PSSE) is: 
\begin{equation} 
\text{\textit{PSSE}}_{\lambda}(x_{i}(t)|y))=\sum_{j=1}^{N} [y_{ij}-x_{i}(t)]^{2}+ \lambda \int (D''(x_{i}(t))^{2}dt, 
\end{equation} 
where $x_{i}(t)=\sum_{j=1}^{p} \gamma_{ij} \psi_{j}(t)$ is the basis expansion of each curve and $y_{ij}=x_{i}(t_{j})$ with $j=1,\dots,N$ are discrete observations for the $i$-th curve. Here the roughness of the curve is measured by its integrated squared second derivative and the smoothing parameter $\lambda$ controls the trade-off between the closeness of fit to the average of the data and the variability of the curve. In practice, it is common to choose smoothing parameter subjectively or select it through generalized cross-validation criterion.
\subsection{Functional Principal Component Analysis}
Functional Principal Component Analysis (FPCA) is the extension of the more classical multivariate PCA to functional data and represents a useful tool for displaying curves into a reduced dimensional space. FPCA is one of the main tools considered when clustering functional data, but it can be also applied for data projection and interpretation. \\
Given $n$ functional observations $x_{i}(t)$, $1\le i \le n$, let $\bar{x}(t)$ be the estimate of the mean function. The estimated covariance function, in analogy with the covariance matrix of the multivariate case, is defined as: 
\begin{equation} 
S(s,t) = \frac{1}{n-1} \sum_{i=1}^{n} (x_{i}(s)-\bar{x}(s))(x_{i}(t)-\bar{x}(t)). 
\end{equation} 
As in the multivariate case, under mild assumptions Mercer's theorem (\cite{Mercer}) leads to a spectral decomposition, providing a countable set of positive eigenvalues $\lambda_{1} \ge \lambda_{2} \ge \dots$ associated to a basis expansion of orthonormal basis functions $\psi_{l}(t)$ with $l=1,\dots$ such that 
\begin{equation} 
S(s,t) = \sum_{l=1}^{\infty} \lambda_{l} \phi_{l}(s) \phi_{l}(t). 
\end{equation} 
In standard terminology, the basis functions $\phi_{l}(t)$ are the eigenfunctions or harmonics. The eigenvalues measure the variability in the directions corresponding to the eingenfunctions. \\
Principal components are zero-mean uncorrelated random variables, defined on the same interval of the functional data, with variance $\lambda_{l}$. Principal component scores of the unit $i$ in the dataset are defined as 
\begin{equation} 
c_{i,l} = \int x_{i}(t) \phi_{l}(t) dt. 
\end{equation} 
With these definitions, the fundamental result of \cite{karhunen1946spektraltheorie} and \cite{loeve1946fonctions} expansion holds and allows to obtain the approximation for a generic curve 
\begin{equation} 
x_{i}(t) = \sum_{l=1}^{\infty} c_{i,l} \phi_{l}(t). 
\end{equation} 
If one consider the first $q$ terms of the decomposition, the expression leads to a possible dimension reduction. The information on the curve $x_{i}(t)$ is then synthesized by the $q$-dimensional vector $c=(c_{i1},\dots,c_{iq})$.
\subsection{Functional cluster analysis}
The infinite dimensionality of functional data constitutes a common problem to all clustering methods and leads to some additional difficulties, like the lack of definition for probability density of a functional random variable, the definition of distances or estimation from noisy data. To overcome these problems several methods have been developed that can be mainly grouped on three approaches: two-stages clustering, distance-based clustering or non parametric clustering and model-based clustering.
\subsubsection{Two-stages approach}
Two-stages approach first reduces data dimension by approximating the curves with a finite number of parameters (filtering step) and then uses clustering algorithms for finite dimensional data (clustering step). Filtering step can be performed either by curves' coefficients in a basis of functions or by their first principal components and classical clustering algorithms can be used on them. From a computational point of view, reduction technique by functional principal component scores also needs a basis expansion of curves. The first contribution to two-stages methods is due to \cite{abraham2003unsupervised}, where \textit{k-means} clustering is based on \textit{B-splines} coeffients.
\subsubsection{Distance-based approach}
Distance-based methods for clustering consist generally in defining specific distances or dissimilarities for functional data and then apply clustering algorithms with a hierarchical or a \textit{k-means} method. Indeed, because of the large (infinite) number of variables in the functional context the use of classical distances is affected by the curse of dimensionality. Moreover, considering distances can become too restrictive, while the use of a semimetric -- instead of a distance, leads to a reduction of functional space and authorize to consider as equal functional objects that are actually different. \\
A semimetric $d$ in a functional space $F$ is defined as an application on $F\times F$ that takes values in $\R_{+}$ such that autosimilarity, symmetry and triangle inequality are fullfilled, but not identity property 
\begin{equation} 
d(X_{i},X_{i'})=0 \Rightarrow X_{i}=X_{i'}, \enskip \forall (X_{i},X_{i'}) \in F \times F. 
\end{equation} \\
The families of semimetrics most widely used are based on derivatives and on principal components (\cite{ferraty2006nonparametric}). Principal components can be used for computing proximities between two curves $X_{i}$ and $X_{i'}$ in a reduced dimensional space, considering a truncated version of their basis expansion. In case of discrete observations $x_{i}$ and $x_{i'}$, the empirical version of the semimetric is 
\begin{equation} 
  \label{eq:semimetric_fca}
d_{q}^{FPCA}(x_{i},x_{i'})=\sqrt{\sum_{k=1}^{q} \bigg{(} \sum_{j=1}^{J} (X_{i}(t_{j})-X_{i'}(t_{j})) [\xi_{k}]_{j} \bigg{)} ^{2}}, 
\end{equation} 
with $q$ the number of principal components. The semimetric corresponds to the distance between $q$-dimensional vectors of principal component scores of $X_{i}$ and $X_{i'}$.
\subsubsection{Model-based approach}
Model-based approach constructs omogeneous clusters by means of a density mixture model and allows to predict the membership of each observation to one of the clusters. Conditionally to the membership at a cluster, the observations are supposed to come from a common distribution with cluster specific parameters. In the finite dimensional setting, the main tool to estimate the model is the multivariate probability density. In the case of functional data the probability density is not defined, so we assume a density probability on the parameters describing the curves. The first model-based clustering method for functional data has been developed by \cite{james2003clustering}. We here describe the clustering method proposed by \cite{bouveyron2011model}. \\
Let $Z=(Z_{1},\dots,Z_{K}) \in \{0,1\}^{K}$ be an unobserved random variable indicating the group membership of $x(t)$: $Z_{k}$ is equal to 1 if $X$ belongs to the $k^{th}$ group and 0 otherwise. The clustering task aims therefore to predict the value $z_{i}=(z_{i1},\dots,z_{iK})$ of $Z$ for each observed curve $x_{i}(t)$. Each curve $x_{i}$ can be summarized by its basis expansion coefficient vector $\gamma_{i}$, whose distribution is assumed to be a mixture of Gaussians with density
\begin{equation}
  \label{eq:gamma}
  p(\gamma)=\sum_{k=1}^{K} \pi_{k} \phi(\gamma;\mu_{k},\Sigma_{k}),
\end{equation}
where $\phi$ is the Gaussian density function and $\pi_{k}=P(Z_{k}=1)$ the prior probability of group k. This model is referred to as the Functional Latent Mixture (FLM) model (\cite{bouveyron2011model}), since it can be reparametrized and represent the curves through their group-specific eigenspace projection. The spectral decomposition of the matrix $\Sigma_{k}$ allows to model and interpret the variance of the data of the $k$th group through the parameters $a_{k1},\dots,a_{kd_{k}}$ and the variance of the noise through parameters $b_{k}$, where $d_{k}$ can be considered as the intrinsic dimension of the latent subspace of the $k$th group ($\text{FLM}_{[a_{kj}b_{k}Q_{k}d_{k}]}$). Differently than two-stages methods, in which the estimation of these parameters is done previously to clustering, the two tasks are performed simultaneously in this approach. The \texttt{funHDDC} algorithm (\cite{bouveyron2014funhddc}) models and clusters the curves through their projections in the group-specific subspaces obtained by performing functional principal component analysis conditionally on the posterior probabilities of belonging to group $k$.
\section{Data}
\label{sec:Data}
We choose data from Human Mortality Database (\cite{HMD}), that ensures a high quality and quantity of data on mortality profiles of many European and some non-European countries for several year. From the 40 countries available we excluded those with too short time series available (Chile, Croatia, Greece, Israel, Slovenia, Korea, and Taiwan) and those with a too limited population size (Luxembourg and Iceland). As for the time period, we chose to consider data from 1960 (after the Second World War and related economic crises) to 2010. Considering that we need to split Germany into East and West, in order to have data back to 1960, we end up with data for 32 countries and 50 years (see Table~\ref{fig:classif}). 
	\begin{table}[t!]
	\caption{Classification in groups of analysed countries.}\label{fig:classif}
	\centering 
	\begin{tabular}{lp{0.5\textwidth}}
	\toprule
	Area					& Countries \\
	\midrule
	\textbf{North EU}		& Denmark, Finland, Norway, Sweden \\
	\textbf{West EU}		& Austria, Belgium, Switzerland, Germany, France, Ireland, The Netherlands, United Kingdom \\	
	\textbf{South EU}		& Italy, Portugal, Spain \\
	\textbf{Center EU}		& Bulgary, Czech Republic, Hungary, Poland, Slovakia \\
	\textbf{Est EU}		& Belarus, Estonia, Latvia, Lithuania, Russia, Ukraina \\
	\textbf{Extra-EU}		& Australia, Canada, Japan, New Zealand, United States of America \\
	\bottomrule
	\end{tabular}
	\end{table}
This means that for each combination of country and year we have a curve of mortality age pattern. Usually age-specific rates are used for mortality analysis, however we chose to use the age distribution of deaths ($d_x$). We do that because one of the most acknowledged transformation of mortality age patterns in developed countries in the past decades is the shifting (see, for instance, cite{VCanudas2008}) of the modal age at death and the compression of deaths above the mode (\cite{Tatcher2010}). More recently, \cite{Zanotto2016} have shown that premature mortality has also evolved in the last years, with different patterns for several countries. All these transformations are much better visible from the age distribution of deaths ($d_x$) than from the age-specific rates ($m_x$) and this explains why recently new models fitting the $d_x$ are emerging (\cite{Mazzuco2018, Basellini2019}). As \cite{Basellini2019} note, mortality rates ($m_x$), survival probabilities ($l_x$) and age distribution of deaths ($d_x$) are complementary functions, and each one can be derived by each of the others. This means they convey the same information, so choosing one or the other does not affect the results of the cluster analysis, but, as said above, using $d_x$ will allow to more easily visualise the transformations of mortality profiles of selected countries.
\section{Results}
\label{sec:Results}
\subsection{From raw data to smooth curves}
\begin{figure}[t!] 
\centering 
\subfloat[][111 knots, $\lambda_{GCV}=3.548$.] {\includegraphics[width=7cm]{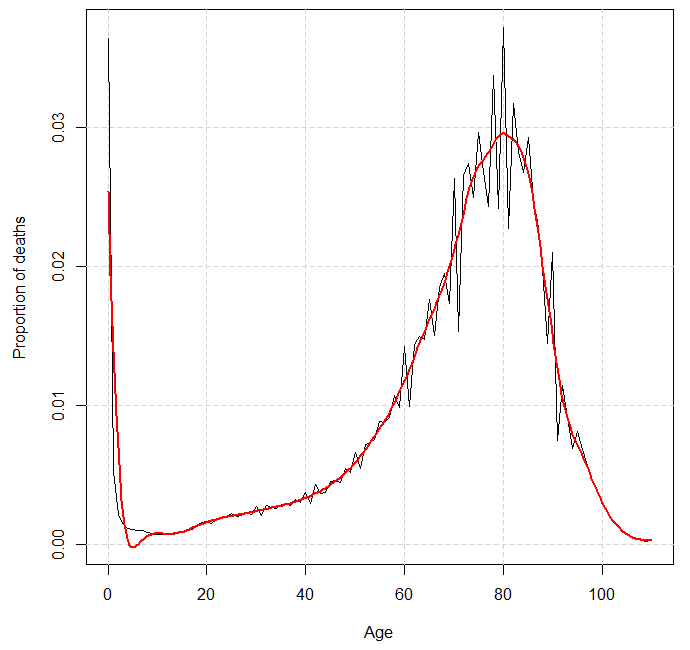}} \quad 
\subfloat[][31 knots, $\lambda_{GCV}=0.016$.] {\includegraphics[width=7cm]{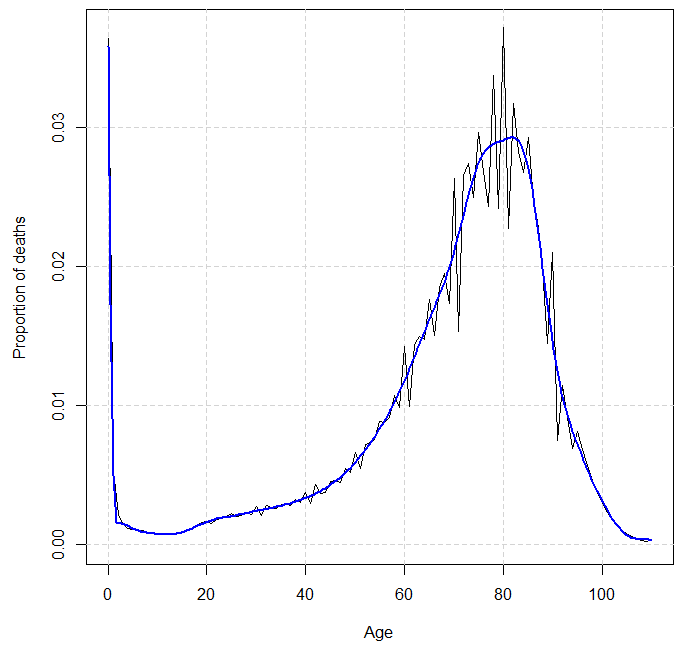}}	
\caption{Smoothing with two sequences of knots for the curve of Russia in 1960.} 
\label{fig:selezione_lambda} 
\end{figure}

The analyses of this section will focus on the study of mortality curves to understand the patterns or trajectories through the selected time period for the principal developed countries. Although in functional analysis there is no general requirement for the data to be smooth, we can find in some cases, particularly noisy data which makes smoothing necessary. This problem affects the most the curves of Eastern countries at the beginning of the time period, and is attributable to quality of data. \\
We used the \textbf{R} package \texttt{fda} of \cite{ramsayfda} to obtain a functional representation using a basis expansion of natural cubic splines. In order to mantain the data structure two sequences of knots over the age range [0,110] have been evaluated: a sequence of 111 equally distributed knots (i.~e. one for every age) and a sequence of 31 knots, one every 3 months over age interval [0,2], one every 5 years over age interval [2,110]. The latter has been preferred to the former, not only as it is more parsimonious, but even because it is preferable in terms of goodness of fit. As an example, in Figure~\ref{fig:selezione_lambda} both solutions of knots sequence are applied for the curve of Russia in 1960. 
The comparison reveals that 31 knots unequally distribued allow to account for the steep decreasing of infant mortality in the first two years and the unicity of the mode distribution. In this example, the smoothing parameters for each of the two sequences of knots have been selected through Generalised Cross-Validation (GCV) criterion ($\lambda_{GCV}=3.548$ with 111 knots, $\lambda_{GCV}= 0.016$ with 31 knots). GCV is mean-squared error based measure, twice discounted by a term taking into account number of parameters and magnitude of smoothing parameter.\\
In the following analyses, considering the curves for all the countries and years, two alternatives for the smoothing parameter have been applied: a common smoothing parameter for all curves ($\lambda_{GCV}^{COM}=0.0025$) and a different smoothing parameter for each curve. As the results for the two alternatives do not show any relevant differences, we present only the ones obtained with a curve specific $\lambda$.
\subsection{Exploratory analysis through FPCA}
\begin{figure}[t!] 
\centering 
\subfloat[][Men.] {\includegraphics[width=14cm]{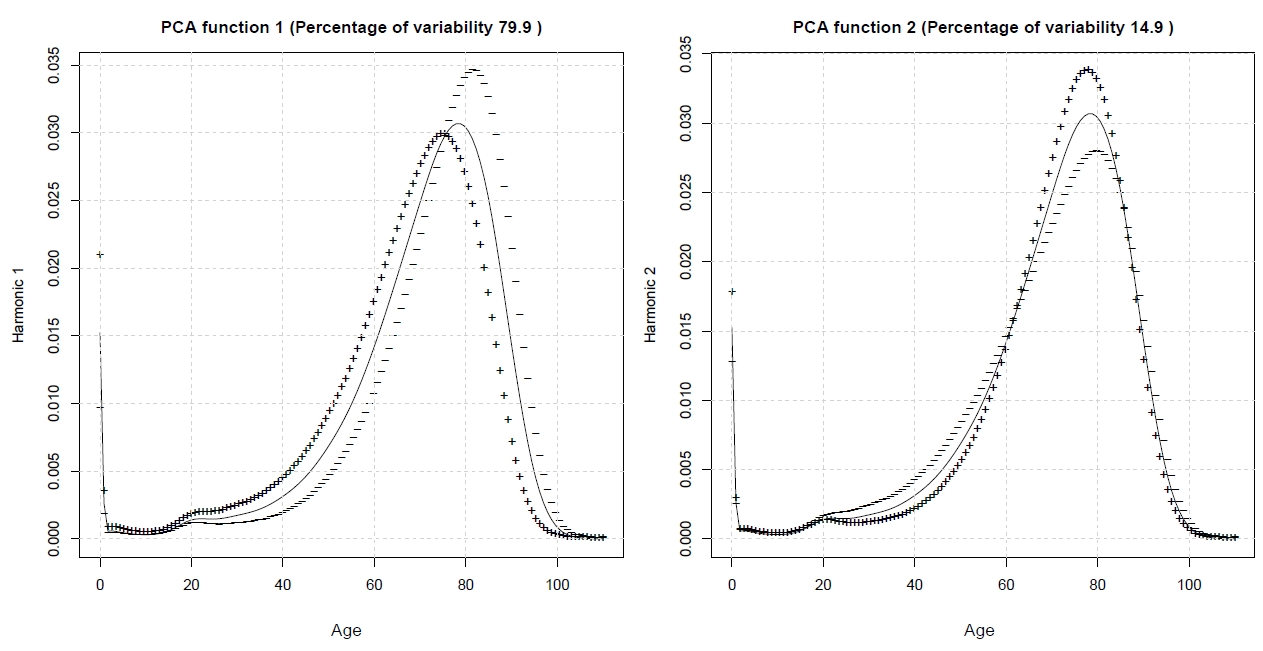}} \quad
\subfloat[][Women.] {\includegraphics[width=14cm]{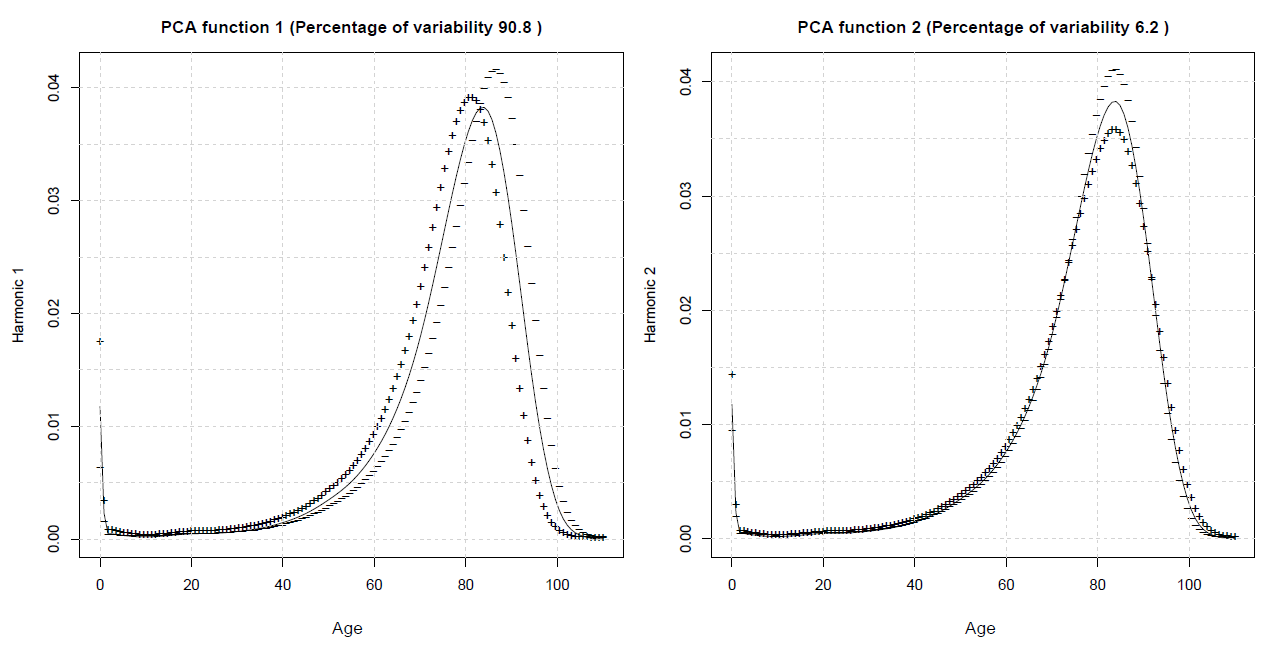}} 
\caption{Effect of the first two principal components on mortality data: the overall mean and the mean $\pm$ a suitable multiple of the principal component weight function.}
\label{fig:components} 
\end{figure}

\begin{figure}[t!]
\centering 
\subfloat[][Men.] {\includegraphics[width=15cm]{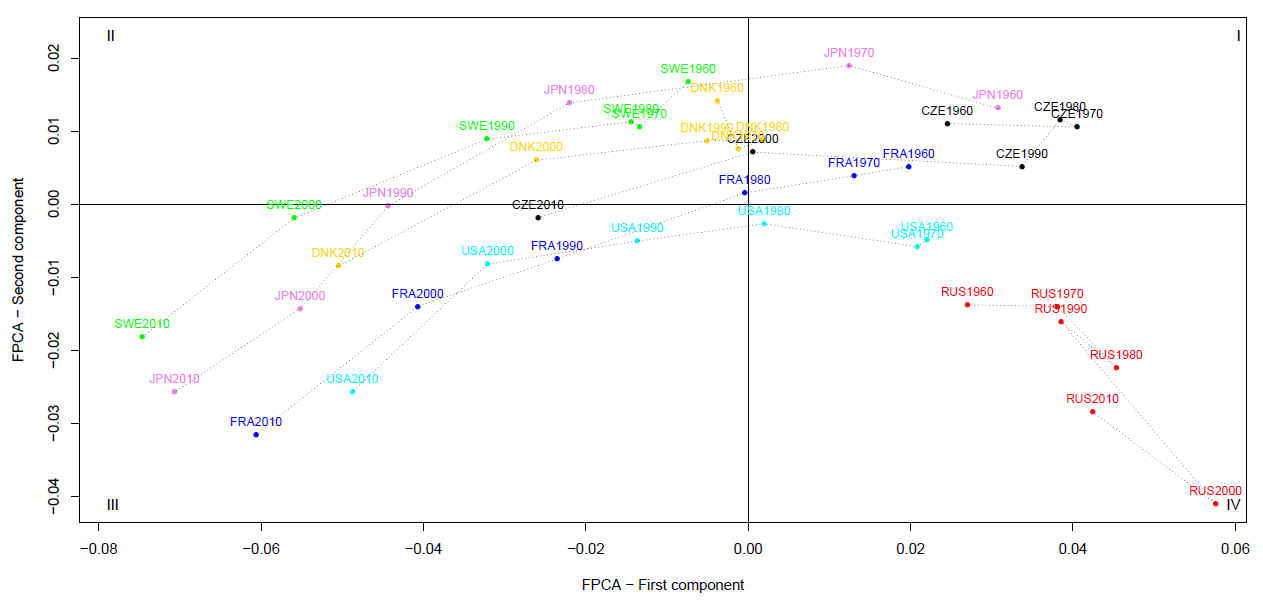}} \quad
\subfloat[][Women.] {\includegraphics[width=15cm]{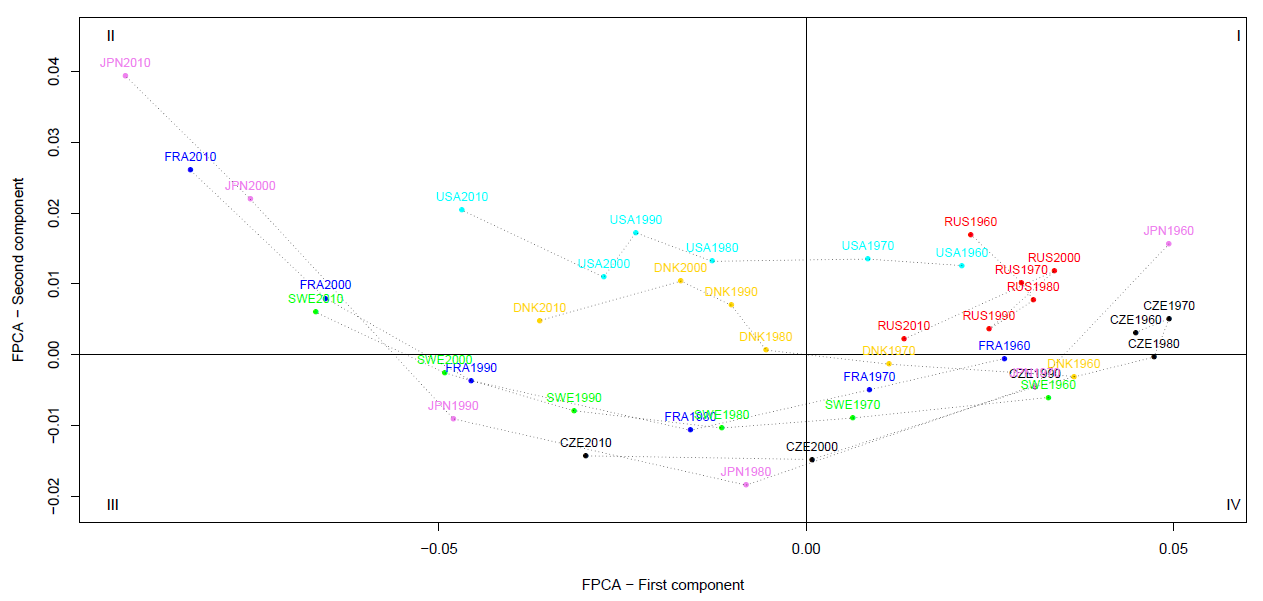}} 
\caption{First principal subspace.}
\label{fig:subspaces} 
\end{figure}

\begin{figure}[h!] 
\centering 
\subfloat[][France in 2010 - Men.] {\includegraphics[width=7cm]{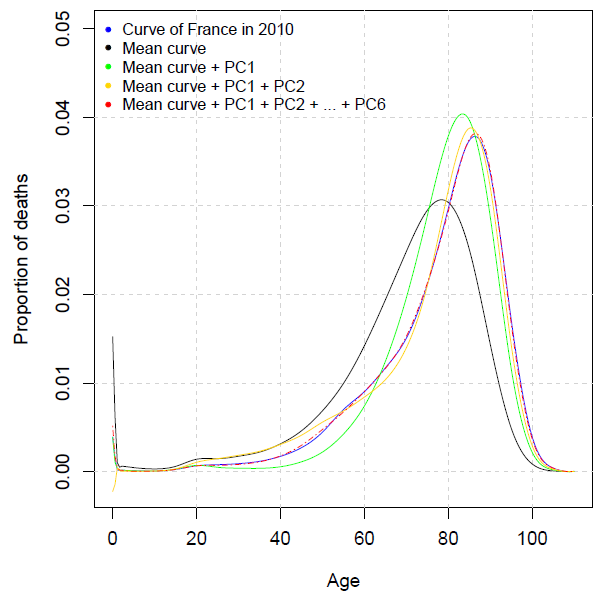}} \quad
\subfloat[][Japan 2010 - Women.] {\includegraphics[width=7cm]{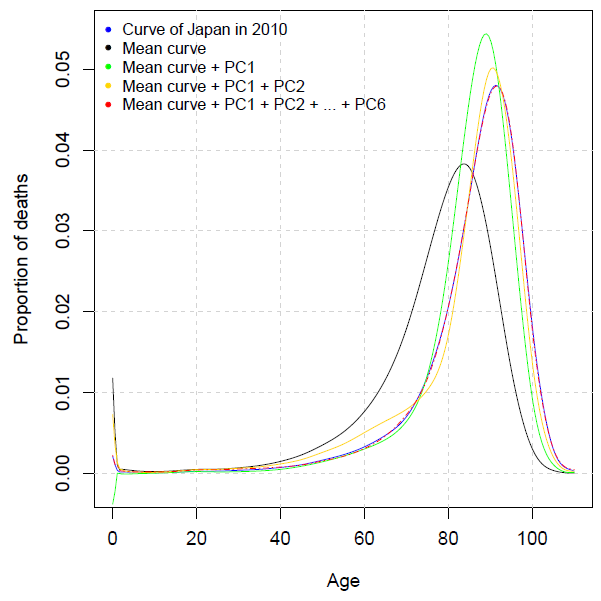}}
\caption{Reconstruction of curves through functional principal components.}
\label{fig:decomposition} 
\end{figure}

The FPCA has been performed in order to synthetize the variability of the curves. Many works on mortality evolution treat separately data for males and females, due to the fact that they experienced in the past different mortality trends. In the same spirit, we represented functional data with curve-specific smoothing parameters for both sexes and conducted two FPCA. It emerged that most of the variability is explained by the first two principal components both for men and for women (95\% for both). A classical way to interpret the principal component functions is to plot the group mean function as well as the functions obtained by adding and subtracting to the mean function twice the square root of the principal component variance ($\bar{X}(t)\pm 2 \cdot \sqrt{\lambda_{i}}\xi_{i}(t)$), with the $\lambda_{i}$ eigenvalue of the $i$th component). Refer to \cite{Ramsay1, ramsay2005analysis} for more details on this usual representation. \\
In Figure~\ref{fig:components} for each of the first two components three curves are plotted: the dashed curve is the overall smoothed mean, which is the same by sex, whereas the other two curves show the effect of adding and subtracting a suitable multiple of the principal component weight function. 
Looking at men (a) one can see that the first component corresponds to a shift of the curves respect to the overall mean of deaths over the entire age range. The (+) curve has an higher mortality of the (-) curve respect to the mean curve before 80 years, lower afterwards. In addition, an increase in the number of deaths is observed around modal age at death. A high scorer on this component woulds show above-average shift. For what concerns the second component, the variability is concentrated between 20 and 60 years and around modal age at death. This variability opposes the (-) and (+) curves, which cross at 65 years approximately. The (-) curve lies above (+) curve beween 20 and 60 years and below (+) curve around modal age at death. A high scorer on this component expresses a low premature mortality between 20 and 60 years. Therefore, we can summarize that the first component is representative of the shift and compression of deaths distributions observed in the latest dacades, while the second component is related to premature mortality. \\
This is already an interesting results, as it confirms that shift and compression of mortality schedules are intertwined (\cite{Bergeron2015}) and it reveals that premature mortality component is independent on shift and compression and 15\% of variability in men mortality schedules are attributable to it. Concerning the women (b), the first component reflects a shift and compression from age 40 throughout adulthood and senescence, weaker compared to men. The second component (6\% of variability) shows an increase in the number of deaths around modal age at death not attribuable to a clear change in premature mortality. \\
The principal subspace allows to give each individual a score in terms of the attributes expressed by principal components. In Figure~\ref{fig:subspaces} we can find the scores of seven representative countries (Denmark, Sweden, Japan, France, Czech Republic, United States, Russia) on the two first components, selected every 10 years over the time period for ease of interpretation (see Appendix A for the plot of all considered countries).
The first principal subspace for men (a) shows similar trajectories on the first component for Denmark, Sweden, Japan, France, Czech Republic, and United States. The first axis discriminates these countries throughout the whole period from the I quarter to the II quarter; the decrease of the scores reflects the shift of mortality curves towards older ages with respect to the mean curve. Even though direction is similar, the pace of evolution differs among countries; Sweden starts in advance compared to the other countries, a delay of two decades can be noticed for Czech Republic and Denmark stagnates between the 70s and 90s. The second axis indicates for Denmark, Sweden, Japan, France, Czech Republic, and United States different levels of premature mortality already from the starting point in the I quarter. United States starts lower than other countries reflecting an higher premature mortality, while the curves of Japan and Sweden are the ones with an higher number of deaths around modal age at death. Again, we can observe the stagnation of Denmark on the second component between the 70s and 90s. Only Russia experiments a completely different trajectory, remaining for the whole period in the IV quarter; the units with the higher premature mortality are the ones of years 1990--2010. Female first principal subspace (b) reports the trajectories in terms of `shift' and `compression', characterized by a shift of mortality curves for Denmark, Sweden, Japan, France, Czech Republic, and United States and a permanency of curves on lower ages for Russia. \\
The decomposition property of Karhunen-Loève turns out useful for the evaluation of the appropriate number of principal components to obtain an exact approximation of the curve. Figure~\ref{fig:decomposition} shows the reconstruction of two smoothed curves (France and Japan in 2010) obtained from the mean curve by adding the principal components one at a time. We can see for men that considering the first two components is not enough and the first six components are needed. As a consequence, one has to be careful also on the interpretation of French increase in male premature mortality in the last years of time period, as suggested by negative scores on the second component (Figure~\ref{fig:subspaces} (a)). In this case, the effect of the second component is over-estimated and reduced by the following four components. 
The difficulty of reconstructing the curve with the first two components could be linked to the change of shape of death distribution. In this respect, \cite{zanotto_twoviews} hypothesize that premature mortality is not increasing because of a specific cause of death, but rather the shift and compression in a reduced range at older ages could have isolated and emphasized it. \\
The reconstruction of the curve of Japan in 2010 (Figure~\ref{fig:decomposition} (b)) suggests another pattern of evolution. Both the principal components express the compression of the curve and the second reduces the effect of the first one. Therefore the high value of second component does not only indicates a decrease in the number of deaths around modal age at death compared to the mean curve, but rather a faster pace for shift than compression. The phenomenon has been described by \cite{VCanudas2008} as a ``shifting mortality scenario, where bulk of deaths around the modal age at death move toward older ages and the compression of mortality has stopped. This may be a realistic description of the current situation in low mortality countries.''
\subsection{Analysis of mortality evolutions}
In this section we present the results of the classification of mortality curves for the 32 countries selected from 1960 to 2010. The three methods of cluster analysis have been carried out, reflecting the three main approaches for functional clustering: two-stages approach on the coefficients of basis expansion of the curves, model-based approach with the FLM model and distance-based approach through semimetric based on FPCA. We chose to show distance-based method for women and model-based method for men. This choice stems from the different pattern shown by men and women mortality cluster solutions and would be explained in the following. Mortality curves and corresponding mean curves within the clusters will be analysed, as well as the composition of clusters in terms of countries and years. Model-based method has been performed with the package \texttt{funHDDC} (\cite{bouveyron2014funhddc}), whereas for distance-based approach through semimetric we used the package \texttt{fda.usc} (\cite{de2011utilities}), which extend the functionalities of \texttt{fda} package (see Endnote \footnote{The R code used is available on github, so results are fully replicable.}).
\subsubsection{Cluster analysis for men}
A cluster analysis has been conducted for men following model-based approach. Since we obtained the data in a common acquisition process, a consequent natural assumption is that the behavior of the error components outside the class specific subspaces is common. Thus we modeled the noise outside the latent subspace of the group by a single parameter $b$ and chose the reduced model $\text{FLM}[a_{kj}b_{}Q_{k}d_{k}]$. The number of clusters has been selected according to BIC criterion, defined with a positive log-likelihood. As a monotone trend for BIC is not verified and the BIC doesn't stabilize, but two local maxima occur at $K=5$ clusters and $K=7$, we show the partition in seven clusters, which allows for more flexibility and better interpretation. Table~\ref{fig:latent_model} reports the informations on the model within the $K=7$ clusters. The dimensionality $d_{k}$ varies between 1 and 5 and the complexity of the model, controlled by $K$ and $d_{k}$, is equal to 843 parameters. The number of parameters $a_{kj}$ ($k=1,\dots,7$, $j=1,\dots,d_{k}$) corresponds to the eigenvalues selected for every $\Sigma_{k}$. The stability of cluster dimensions has been verified by initializing the classes of \texttt{funHDDC} algorithm with \textit{k-means} function and setting different seeds. \\
The partition in 7 clusters (Figure~\ref{fig:model-based}) identifies the curves with high infant mortality and the shift towards older ages; furthermore, it groups less compressed curves and those with higher premature mortality. Cluster 1 contains the curves with high premature mortality (4\% on average) and cluster 3 the ones with a similar shape but lower infant mortality (2\% on average). Cluster 2 expresses the increase in premature mortality and a stronger decrease of the number of deaths around modal age at death. The curves in cluster 4 are more compressed and the number of deaths is lower around modal age at death. Cluster 5, 6, and 7 show continuous shift and compression of mortality curves.
	\begin{table}[t!]
	\caption{Model $\text{FLM}[a_{kj}b_{}Q_{k}d_{k}]$ ($b=0.008$, Cattel scree test threshold = 0.2). 
		Values of parameters have been magnified by 10,000.}\label{fig:latent_model}
	\centering 
	\begin{tabular}{lccccccc}
	\toprule
	& $d_{k}$		& $a_{1}$ & $a_{2}$ & $a_{3}$ & $a_{4}$	& $a_{5}$	 \\
	\midrule
	Cluster 1 & $3$	& $0.95$ 	& $0.59$ 	& $0.29$ 	& $$ 		& $$	 \\
	Cluster 2 & $1$	& $1.70$ 	& $$ & $$	& $$ 		& $$		& $$ \\
	Cluster 3 & $5$	& $0.53$ 	& $0.37$ 	& $0.24$ 	& $0.085$	& $0.0054$	 \\
	Cluster 4 & $4$	& $0.56$ 	& $0.27$ 	& $0.14$ 	& $0.11$ 	& $$	 \\
	Cluster 5 & $2$	& $0.54$ 	& $0.39$ 	& $$ 		& $$ 		& $$	 \\
	Cluster 6 & $2$	& $0.71$ 	& $0.35$ 	& $$ 		& $$ 		& $$ 	\\
	Cluster 7 & $2$	& $1.20$ 	& $0.37$ 	& $$ 		& $$ 		& $$	\\
	\bottomrule
	\end{tabular}
	\end{table}
\\
We summarize the features of the evolution emerged from the analyses of the six areas, highlighting the specificities of some countries. Northern, Western, Southern, and extra-European countries experiment a shift of the curves and an increase in the number of deaths around modal age at death over the whole period (cluster 4, 5, 6, 7). Norway, Sweden, and Netherlands, known for their high values of longevity, are in advance from the beginning of the time period (starting already from cluster 5). Among Western countries, Switzerland and France anticipate the evolution process (first ones passing to cluster 7), while Eastern Germany and Ireland evolve slower than other countries. The analyses also identify the delay of Finland in the first twenty years and the stagnation Denmark between 70's and 90's, which lays behind all countries at the end of time period (last passing to cluster 7). Southern countries show the highest infant mortality in the early twenty years (cluster 1) and then follow the shifting and compression process of mortality curves already described (cluster 5, 6, 7). Among extra-European countries, United States are characterized, for the first two decades, by less compressed curves with respect to other countries (cluster 3). The rapid shift of Japan from the second half of the period is coherent with the strongest increase recorded in longevity (already in cluster 7 in 1985). \\
For Central countries one can observe an high infant mortality in the first decade of the period (cluster 1), made exception for Czech Republic. Then, their evolutions differ considerably. Czech Republic evolve similarly to Western countries but cumulates a delay of about twenty years with respect to them (reaching cluster 6 in 2005). In Hungary premature mortality starts to increase from the 80's and continue until 2010 (cluster 4), with no sign of reversing. Bulgaria, Poland, and Slovakia stop compression in the 80's and 90's (cluster 3) and show a slight shift in the last decade (cluster 5). The greatest difference between curves concern Eastern Europe. Countries of this area are characterized until the end of 80's by less compressed curves with respect to the other areas (cluster 3) and after the dissolution of USSR by an increase in premature mortality (cluster 2). 
\begin{figure}[t!]
\centering
\includegraphics[width=14cm]{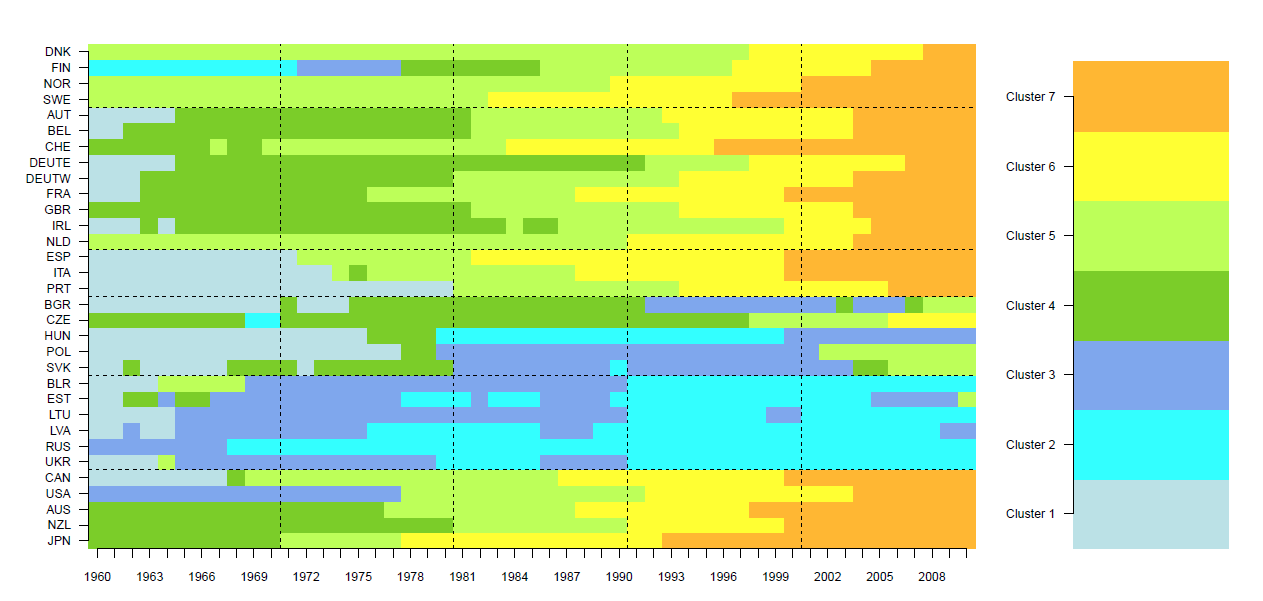}
\includegraphics[width=14cm]{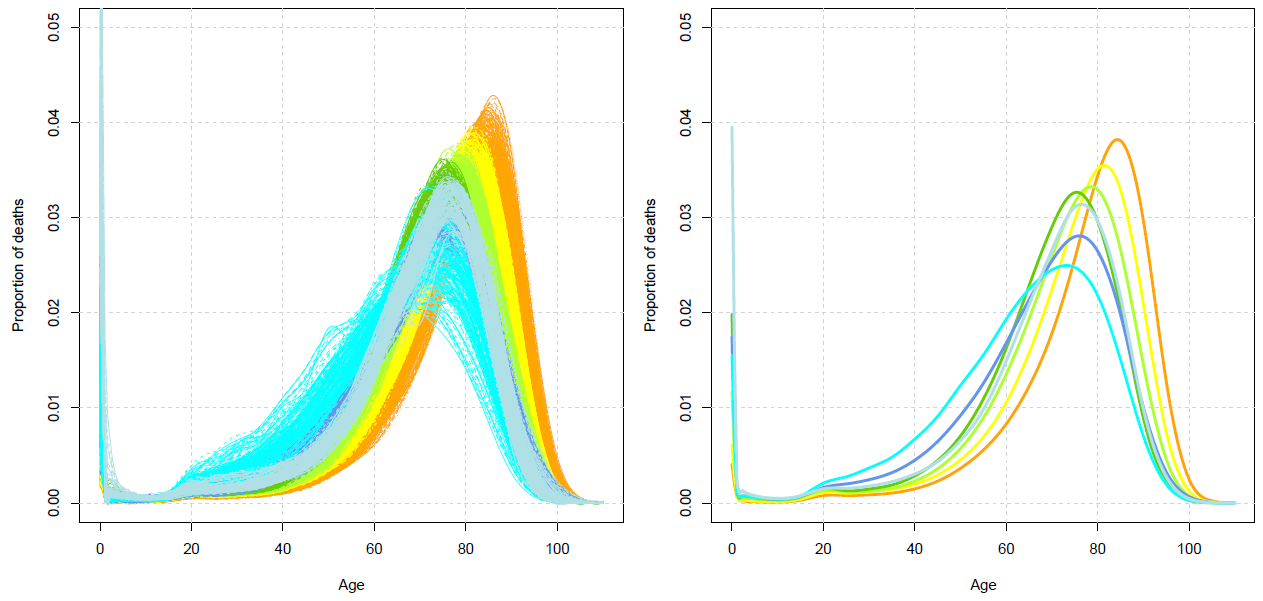}
\caption{Model-based cluster analysis for men - Mortality curves, mean curves and composition of the 7 clusters
	(9.31\%, 11.89\%, 12.87\%, 17.65\%, 21.81\%, 14.77\%, 11.7\% of the units).} 
\label{fig:model-based}
\end{figure}
\\
The other two methods, two-stages and distance-based, identify the evolutions based on the same components of mortality with a lower number of clusters (see Appendix A). The advantage of model-based method consists in the possibility of analyzing the functional subspaces in which data are modeled and classified. As usual, we can plot for each cluster and each selected component the group mean function and the effect of component's variance ($\bar{X}(t)\pm 2 \cdot \sqrt{a_{kj}{\color{white}^{2}}}\xi_{i}(t)$, with the $a_{kj}$ eigenvalue of the $i$th component). This representation is shown for clusters 2, 3, 5, 7 (Figure~\ref{fig:model-based}), since they highlight interesting features. \\
Cluster 2 of Eastern countries after 1990 with high premature mortality is still characterized by an high local variability due to large differences in the departure of curves from the mean. Cluster 3 appears to captures both higher premature mortality of the first half period of Estern countries and mortality compression of the second half period of Central Europe countries, implying an higher dimensionality with respect to the other clusters. For what concerns cluster 5, the variability is explained by shift towards older ages and compression around modal age at death, in coherence with the trend already described for Northern, Western, Southern, and extra-European countries. Cluster 7 express the phenomena highlighted for France by principal component subspaces and the reconstruction of the curves: the shift without compression leading to a stop in the increase of number of deaths around modal age (first component) and the change of shape on the right side of the curve between ages 40 and 60 (component 2). To conclude, the analysis of variability within clusters reveals it does not exists a common dimension expressing a common evolution. However an increasing cumulated explained variability can be noticed from cluster 5 (77\%) to 6 (87\%) and 7 (93\%), reflecting a certain convergence of evolutions for Norther, Western, Southern, and extra-European countries.
\begin{figure}[p!]
\centering
\subfloat[][Cluster 2.] {\includegraphics[width=14cm]{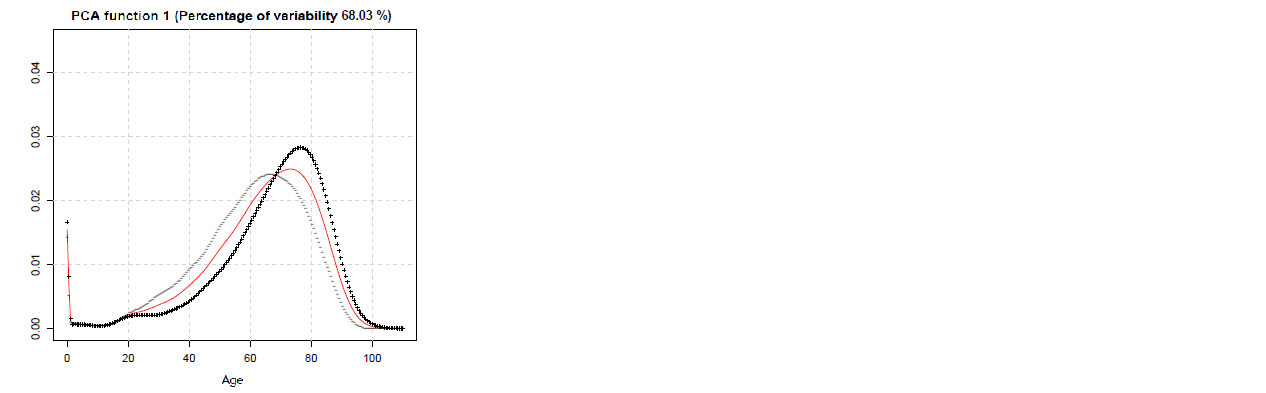}} \\
\subfloat[][Cluster 3.] {\includegraphics[width=14cm]{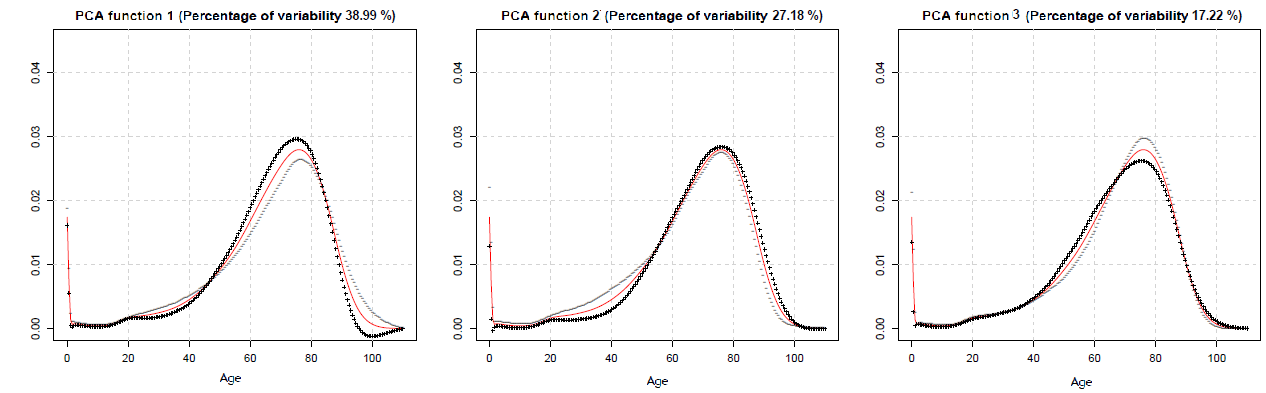}} \\
\subfloat[][Cluster 5.] {\includegraphics[width=14cm]{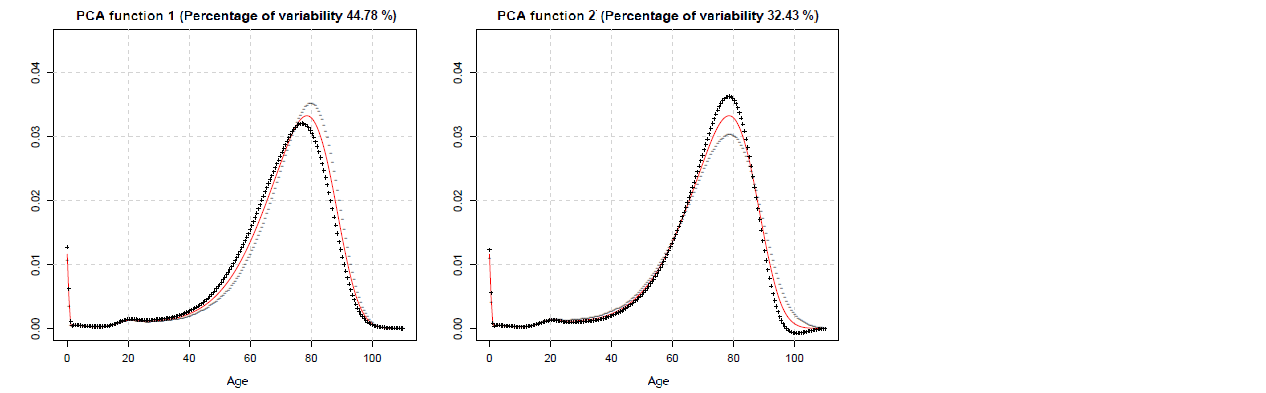}} \\
\subfloat[][Cluster 7.] {\includegraphics[width=14cm]{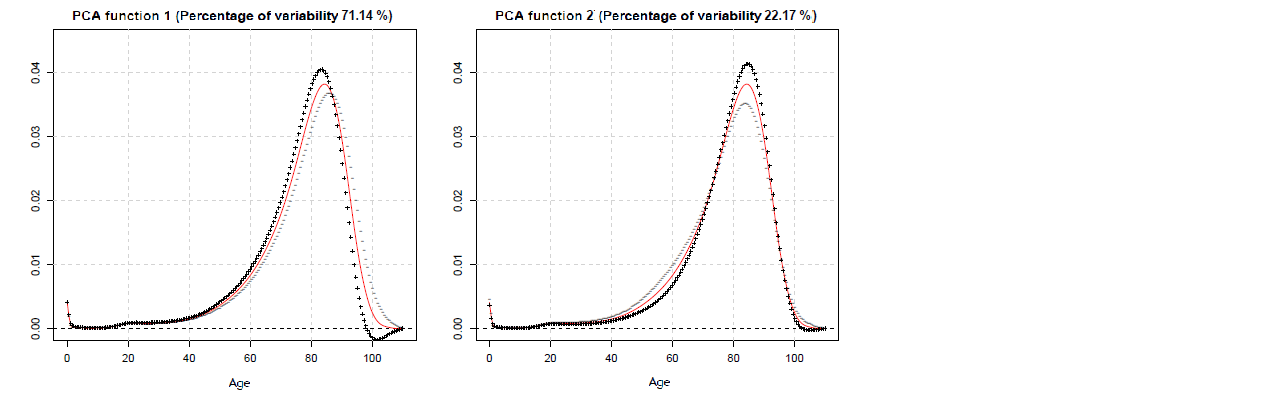}}
\caption{Model $\text{FLM}[a_{kj}b_{}Q_{k}d_{k}]$ for men - Effects of FPCs within clusters.} 
\label{fig:model-based}
\end{figure}
\subsubsection{Cluster analysis for women}
\begin{figure}[t!]
\centering
\includegraphics[width=14cm]{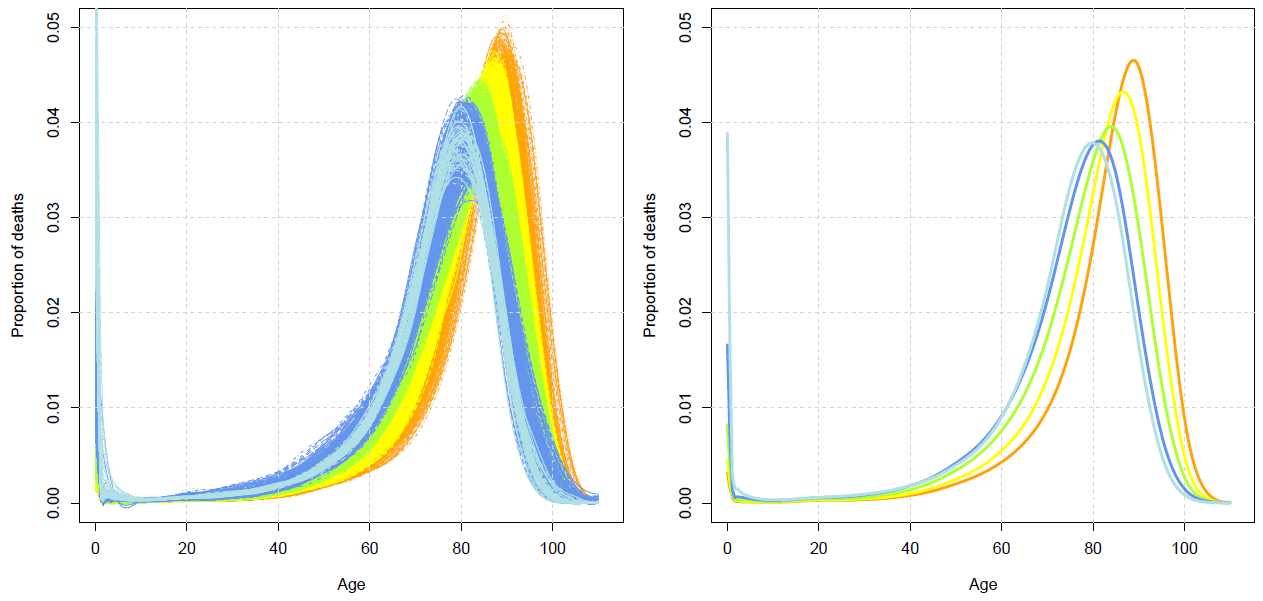}
\includegraphics[width=14cm]{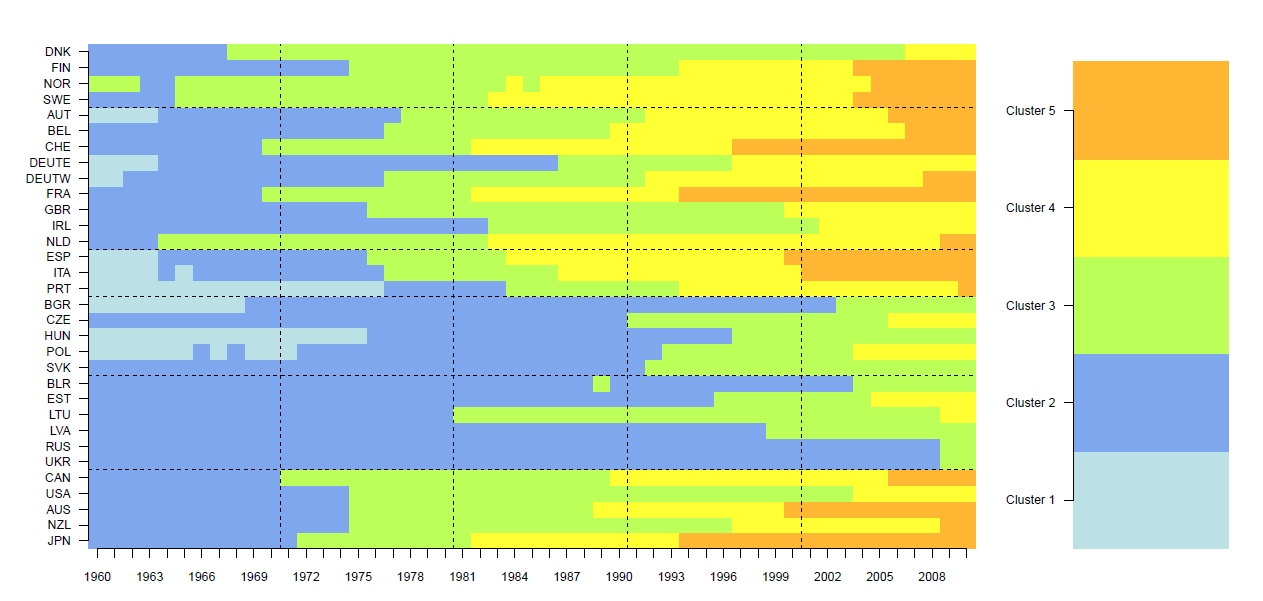}
\caption{Distance-based cluster analysis for women - Mortality curves, mean curves and composition of the 5 clusters
	(4.35\%, 39.15\%, 29.84\%, 19.18\%, 7.48\% of the units).} 
\label{fig:distance-based}
\end{figure}

A hierarchical cluster analysis has been performed for women according to distance-based approach with the semimetric based on the first four FPC (Figure~\ref{fig:distance-based}). Our decision to keep four components is due to the necessity of an approximation of the curves accounting for infant and premature mortality. The number of clusters has been chosen identifying the changes on curves regarding the different components of mortality. In particular, the partition in five clusters allows to disinguish the decrease in infant mortality, the shift of the curves to the right and the increase in the number of deaths around modal age at death. \\
As we can see from Figure~\ref{fig:distance-based}, cluster 1 contains the curves with high infant mortality (4\% on average), in cluster 2 the curves have the same shape but lower infant mortality (2\% on average). Cluster 3, 4, and 5 identifys the curves characterized by a shift to the right and a compression around modal age at death.
\\
The curves with high infant mortality (cluster 1) corresponds to the first decade of Southern countries and some Central countries (Bulgary, Hungary, and Poland). Portugal, Hungary, and Poland mantain an high infant mortality for a longer period compared to other countries (until mid of 70's). The curves of Northern, Western, Southern, and extra-European countries experiment over time period a shift towards older ages (clusters 2, 3, 4, 5). The countries anticipating the shifthing process are Norway, Sweden, and Netherlands at the beginning of 70's, Sweden, Switzerland, France, Spain, and Japan at the beginning of 90's. Denmark lags far behind over the second part of time period and is the last passing to cluster 5 in 2004. The countries of Central Europe and Eastern Europe experiment a long stationary period (cluster 2) and a shift of the curves to the right during the last decade. Although the shift is similar to the one of previous areas, it occur at a slower pace with a delay of about twenty years. Czech Republic, Poland, and Baltic countries seems to be in slight advance (passing to cluster 3 and 4). \\
In the case of women, within-cluster variability does not convey any additional information, as premature mortality is much less important, so model-based analysis does not give any additional insight with respect to distance-based one. The other two methods, two-stages and model-based, identify the evolutions based on the same components of mortality (see Appendix A), even the shift to older ages in the former is emphasizes in the first half period, in the latter in the second half period. \\
To sum up, the analysis for men and women show similar evolutions for Northern, Western, Southern, and extra-European countries, characterized by the shift of curves to older ages and by the concentration of adult mortality around modal age at death. For these four areas we can thereby conclude the existence of a common pattern of evolution. In the case of men all the countries belong to the same group at the end of time period, supporting the hypothesis of an increasing homogeneity. The situation is more heterogeneous for Central and Eastern countries since they don't experiment the same evolution and at the end of time period they don't come to the same cluster. \\
The comparison of the analyses for men and women revealed two different scenarios for Eastern countries, characterizing the increase of premature mortality after 1990 is entirely a male phenomenon. In addition, the shift of the curves towards older ages for Baltic countries is detected only for women. Other differences can be noticed between the two sexes looking at the single countries. The stagnation of Denmark is more pronounced for women in line with \cite{lindahl2016did}, that attribute the stagnation of life expectancy beween 1977 and 1995 to a worsening of health conditions of cohort of women born in the interwar period, linked to smoking behaviour.
\section{Concluding Remarks}
\label{sec:Concl}
In this paper, we have inspected the evolution of mortality schedules in HMD countries by means of a functional clustering method, which allows us to consider mortality patterns as functions, avoiding analysing only a component of mortality (e.~g. infant mortality or old-age mortality) or a summary measure like life expectancy, which is a mixture of all mortality components but without a clear distinction of their contribution to longevity progresses.\\
Three different method of functional clustering have been considered: a principal components based method (FPCA), a distance-based method and a model-based one. The latter method has provided the best results in terms of fit with real data, but FPCA has been also useful in determining what are the most relevant components that drove the transformations we observed in the last sixty years in HMD countries. It turned out that two components account for 94\% of variability in mortality schedules considered: 84\% for a component that can be explained in terms of shifting and compression of mortality, and 10\% for a second component accounting for premature mortality. This demonstrates that shift and compression processes are mutually dependent, while premature mortality is an additional independent component, which accounts for a much lower (10\%) but not irrelevant share of variability.\\ 
The results from clustering also provides us with many insights, although none of them comes as surprise. First the results confirm that an homogeneisation is taking place among most of the considered countries, as many of them follow the same evolution through the clusters. However, men and women patterns are quite different, since for men most of the countries are incuded in the same cluster in the latest years and countries from Eastern Europe not only lag behind with respect to cluster 7, but also do not show signs of a recovering process. Women situation is a bit different because homogeneity of Northern, Western, Southern Europe countries and extra-European ones is less pronounced (Denmark, United Kingdom, United States, East Germany did not reach the highest longevity cluster) but Central and Eastern Europe countries look much closer to precursors.\\ 
The difference between men and women is also characterised by the higher importance that premature mortality has for the formers, so if Eastern Europe countries men longevity is still stagnating, that is partly attributable to premature mortality, notably high in that area. Results also show clearly the stagnation period that Denmark and United States underwent in different periods, much more visible for women. Such a stagnation prevented these two countries to join the highest longevity group. Considering the latest evolution of United States longevity (\cite{JAMA2019}), such lag is going to persist (or even increase) for this countries, while Denmark seems be catching up, as it also can be seen from Figure~\ref{fig:subspaces}. \\
This work, however, was also meant to show the potentialities of Functional Data Analysis demographic studies, where the leading forces of population growth (fertility, mortality and migration) are often measured in terms of age-specific rates or probabilities, that reveals several components. So similar analyses can be implemented on fertility and migration age patterns. Moreover, FDA allows other kind of analyses: regression (both on scalar and functional covariates) and hypothesis test. Therefore, we advocate an increasing implementation of such approach to population studies.

\section*{Acknowledgments}
Stefano Mazzuco acknowledges the support from MIUR--PRIN 2017 project number 20177BR-JXS.
\bibliographystyle{apalike}

\bibliography{FUNCLUS}
\newpage

\section*{Appendix A}
\label{sec:app}

	\begin{figure}[h!] 
	\centering 
	\includegraphics[width=15cm]{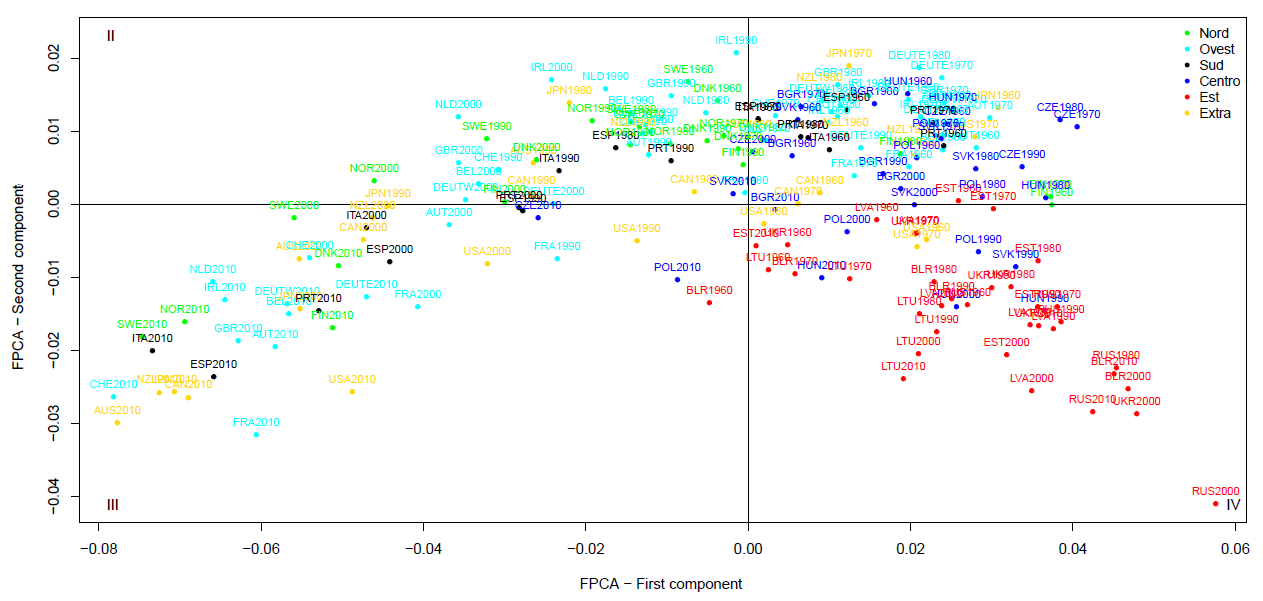}
	\caption{First principal subspace - Men.} 
	\end{figure}
	\begin{figure}[h!] 
	\centering
	\includegraphics[width=15cm]{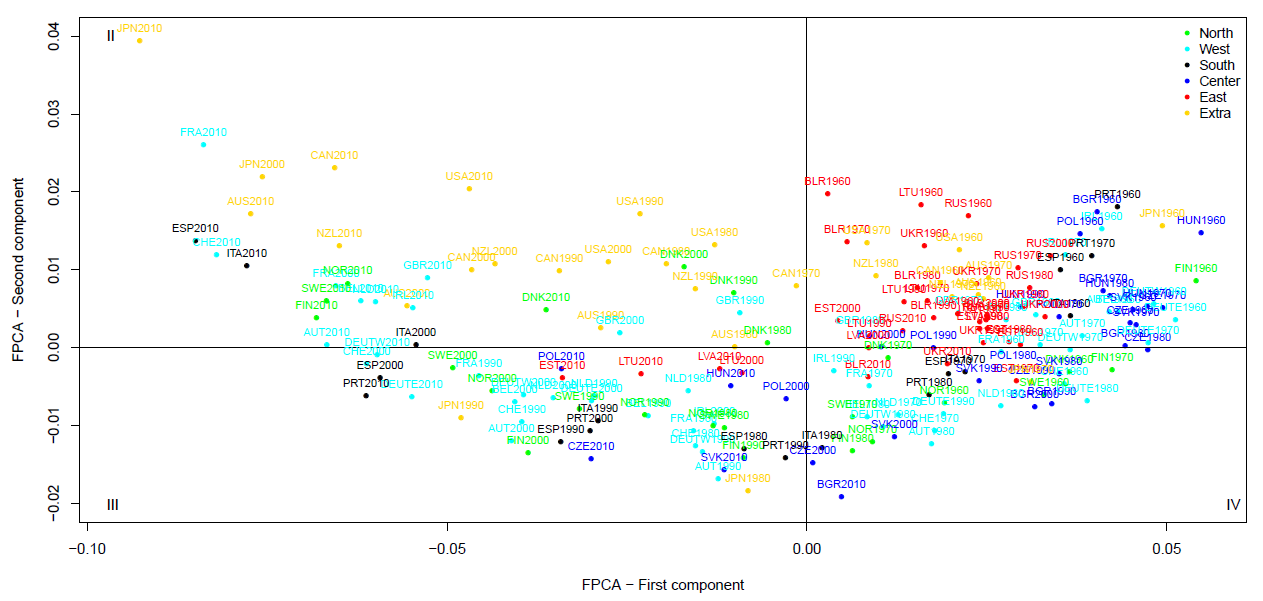}
	\caption{First principal subspace - Women.} 
	\end{figure}

	\begin{figure}[h!]
	\centering
	\includegraphics[width=10cm]{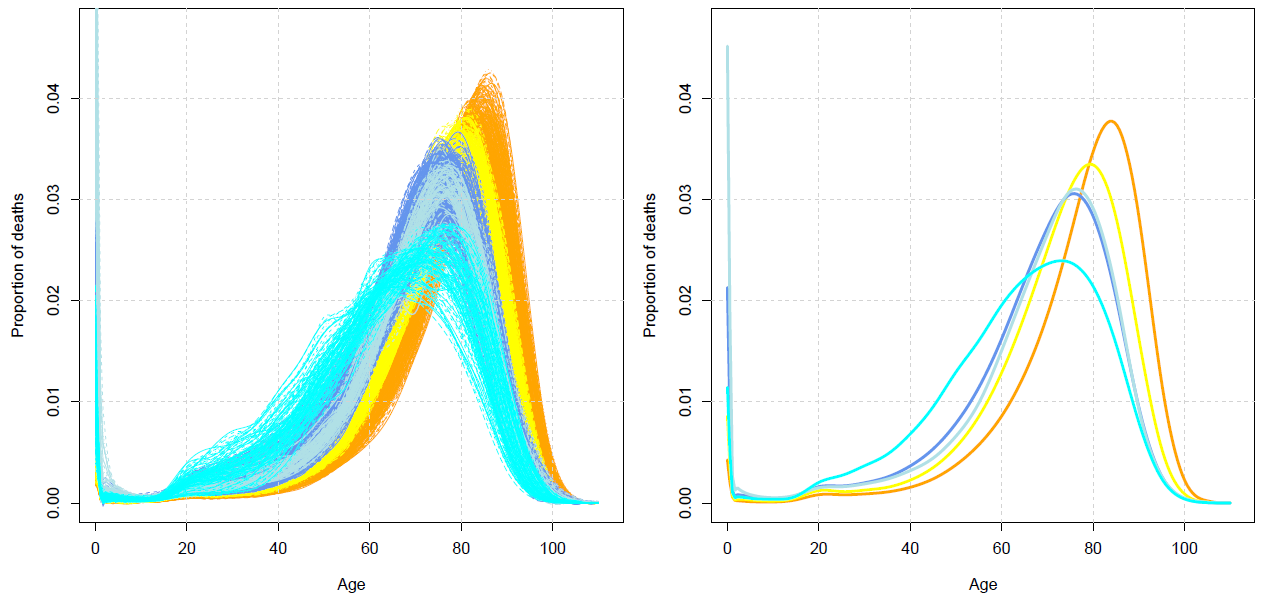} \\[-2pt]
	\includegraphics[width=10cm]{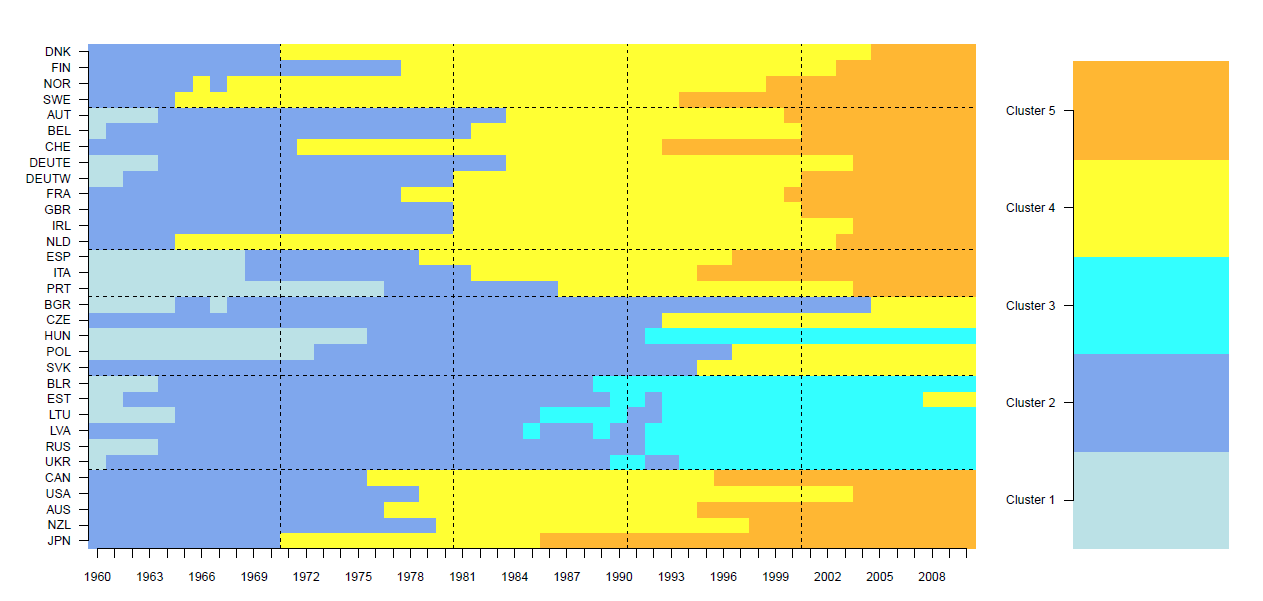}
	\caption{Two-stages cluster analysis - Men.} 
	\end{figure}
	\begin{figure}[h!]
	\centering
	\includegraphics[width=10cm]{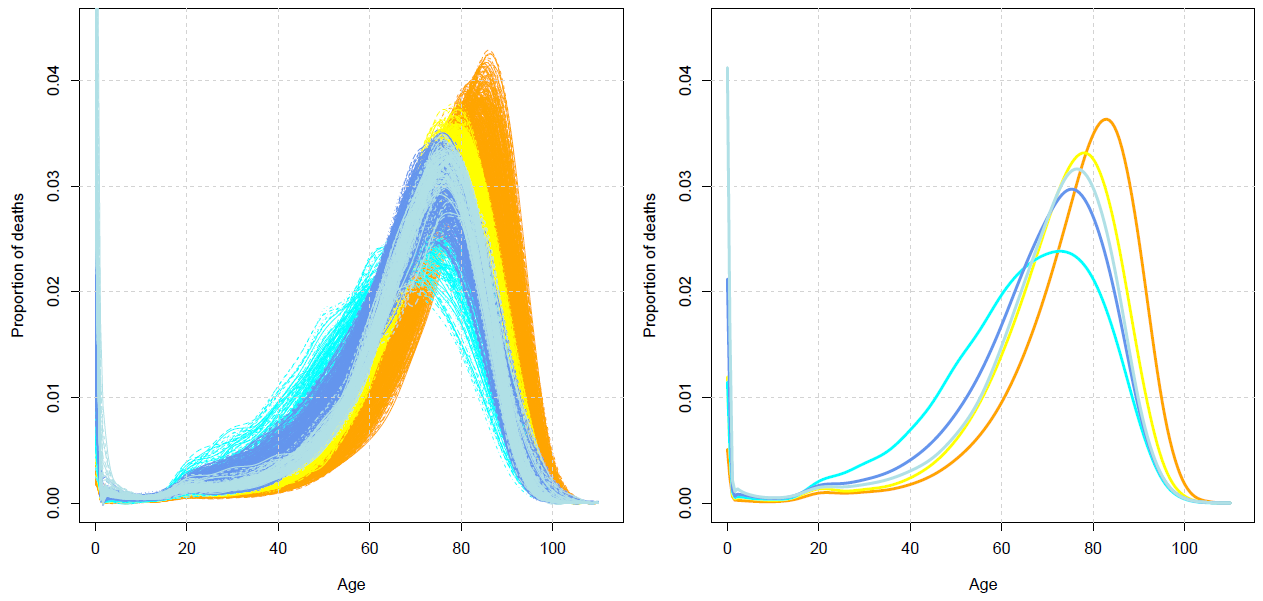} \\[-2pt]
	\includegraphics[width=10cm]{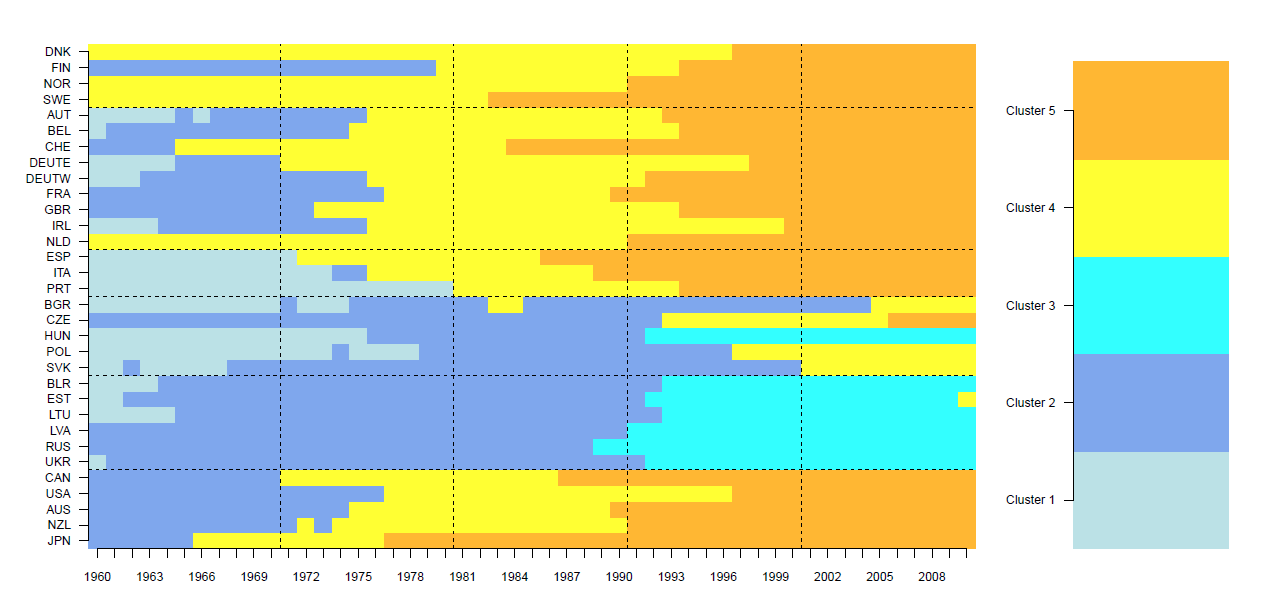}
	\caption{Distance-based cluster analysis - Men.} 
	\end{figure}
	\begin{figure}[h!]
	\centering
	\includegraphics[width=10cm]{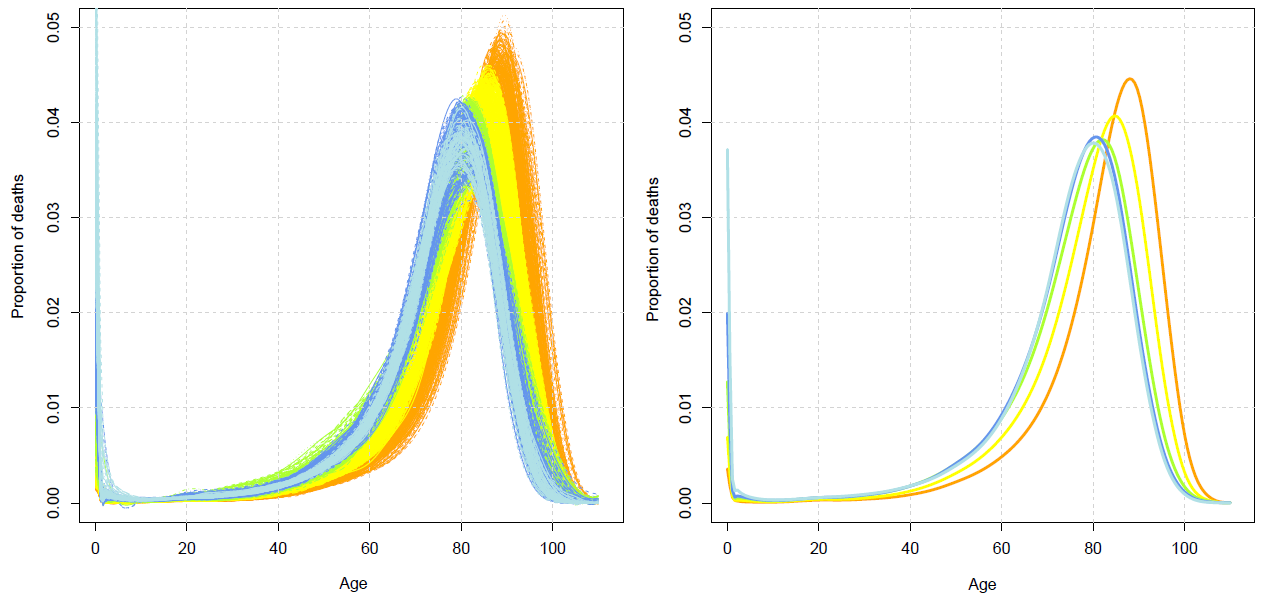} \\[-2pt]
	\includegraphics[width=10cm]{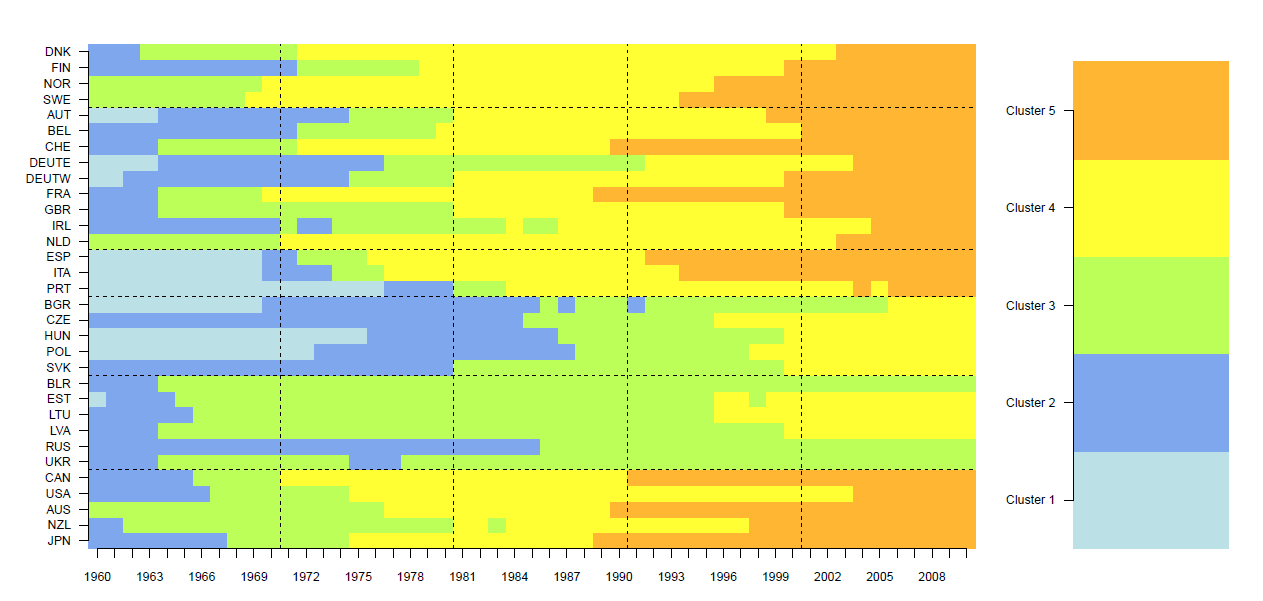}
	\caption{Two-stages cluster analysis - Women.} 
	\end{figure}
	\begin{figure}[h!]
	\centering
	\includegraphics[width=10cm]{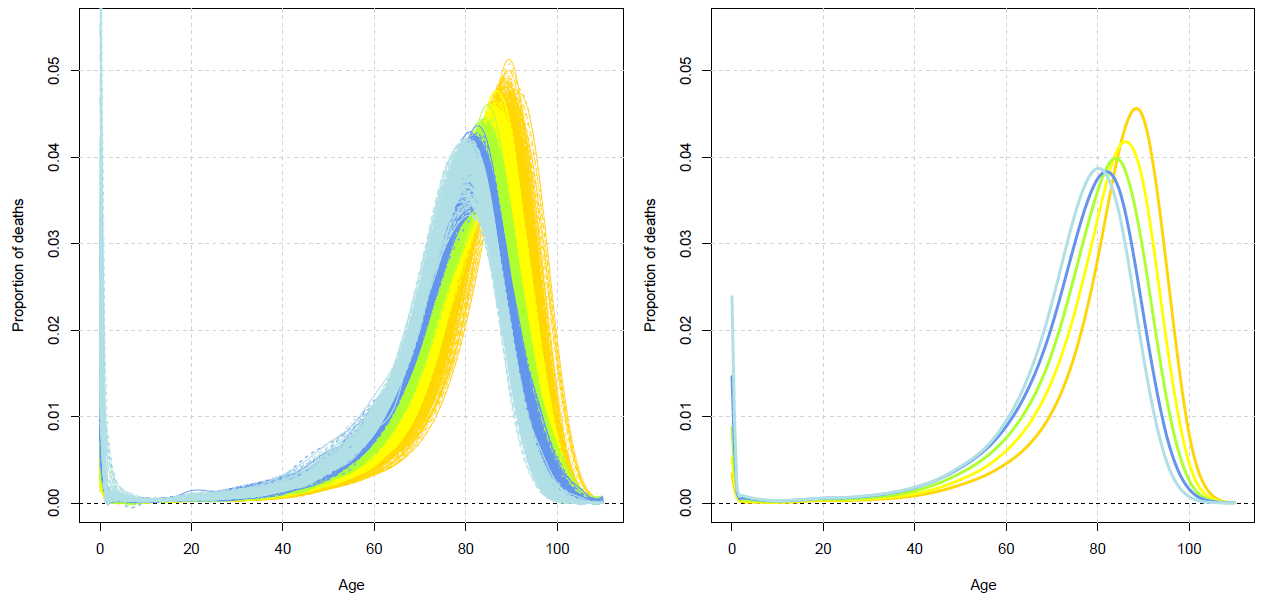} \\[-2pt]
	\includegraphics[width=10cm]{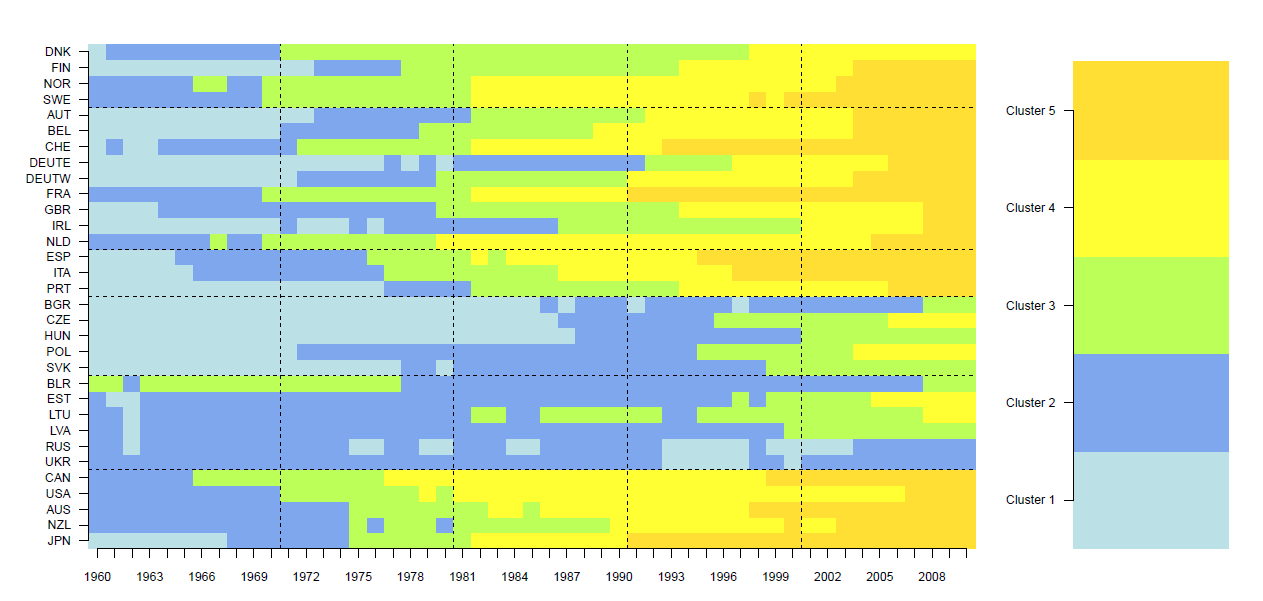}
	\caption{Model-based cluster analysis - Women.} 
	\end{figure}

\end{document}